%% file: main.tex
\documentclass[amsmath,trackchanges,twocolumn,twocolappendix]{aastex702}

\newcommand{\id}{beacon\_1420-5253\_4770}

\def\sfrd{\,{\rm M_\odot\,yr^{-1}\,Mpc^{-3}}}

\usepackage{comment}

\begin{document}

\title{A Type Ia Supernova Candidate at $z\sim4.3$: A Transient Interloper in the Search for $z\sim14$ Galaxies}

\author[0009-0002-5156-7819]{Seiji Toshikage}
\affiliation{Astronomical Institute, Graduate School of Science, Tohoku University, Sendai, Miyagi 980-8578, Japan}
\email[show]{seiji.toshikage@astr.tohoku.ac.jp}  

\author[0000-0002-8512-1404]{Takahiro Morishita}
\affiliation{Astronomical Institute, Graduate School of Science, Tohoku University, Sendai, Miyagi 980-8578, Japan}
\email{morishita@astr.tohoku.ac.jp}

\author[0000-0001-8253-6850]{Masaomi Tanaka}
\affiliation{Astronomical Institute, Graduate School of Science, Tohoku University, Sendai, Miyagi 980-8578, Japan}
\affiliation{Division for the Establishment of Frontier Sciences, Organization for Advanced Studies, Tohoku University, Sendai, Miyagi 980-8577, Japan}
\email{masaomi.tanaka@astr.tohoku.ac.jp}

\author[0000-0003-4299-8799]{Kazumi Kashiyama}
\affiliation{Astronomical Institute, Graduate School of Science, Tohoku University, Sendai, Miyagi 980-8578, Japan}
\email{kashiyama@astr.tohoku.ac.jp}

\author[0000-0002-3407-1785]{Charlotte A. Mason}
\affiliation{Cosmic Dawn Center (DAWN), Denmark}
\affiliation{Niels Bohr Institute, University of Copenhagen, Jagtvej 128, DK-2200 Copenhagen N, Denmark}
\email{charlotte.mason@nbi.ku.dk}

\author[0000-0002-8651-9879]{Andrew J.\ Bunker}
\affiliation{Department of Physics, University of Oxford, Denys Wilkinson Building, Keble Road, Oxford OX1 3RH, UK}
\email{Andy.Bunker@physics.ox.ac.uk}

\author[0000-0001-8587-218X]{Matthew J. Hayes}
\affiliation{Stockholm University, Department of Astronomy and Oskar Klein Centre for Cosmoparticle Physics, AlbaNova University Centre, SE-10691, Stockholm, Sweden}
\email{matthew.hayes@astro.su.se}

\author[0000-0003-3367-3415]{George Helou} 
\affiliation{IPAC, California Institute of Technology, MC 314-6, 1200 E. California Boulevard, Pasadena, CA 91125, USA}
\email{gxh@ipac.caltech.edu}

\author[0000-0002-2993-1576]{Tadayuki Kodama}
\affiliation{Astronomical Institute, Graduate School of Science, Tohoku University, Sendai, Miyagi 980-8578, Japan}
\email{kodama@astr.tohoku.ac.jp}

\author[0009-0005-9953-433X]{Kimi C. Kreilgaard}
\affiliation{Cosmic Dawn Center (DAWN), Denmark}
\affiliation{Niels Bohr Institute, University of Copenhagen, Jagtvej 128, DK-2200 Copenhagen N, Denmark}
\email{kimi.cardoso.kreilgaard@nbi.ku.dk}

\author[0000-0001-9935-6047]{Massimo Stiavelli}
\affiliation{Space Telescope Science Institute, 3700 San Martin Drive, Baltimore, MD 21218, USA}
\affiliation{The William H. Miller III, Dept. of Physics \& Astronomy, Johns Hopkins University, Baltimore, MD 21218, USA}
\email{mstiavel@stsci.edu}

\author[0000-0002-8460-0390]{Tommaso Treu}
\affiliation{Department of Physics and Astronomy, University of California, Los Angeles, 430 Portola Plaza, Los Angeles, CA 90095, USA}
\email{tt@astro.ucla.edu}


\begin{abstract}
The James Webb Space Telescope (JWST) is opening a new window into the distant Universe by discovering galaxies and transients in the early Universe. 
We investigate a high-redshift transient candidate, \id. This object was initially identified as a high-redshift galaxy candidate at $z\sim14$. 
However, the source was not detected in epochs before and after the detection epoch, suggesting that the object is a transient source rather than a persistent galaxy.
We classify the source by comparing the colors, magnitudes, light curves, and spectral energy distribution with various spectral templates of transients. 
Our analysis shows that the observed properties are consistent with a Type Ia supernova at $z\sim4.3$. 
Strong absorption by Fe-group elements seen in Type Ia supernova spectra can mimic the Lyman break used to detect high-redshift galaxies.
At $z \sim 4.3$, corresponding to a cosmic age of only $\sim 1.5$ Gyr, our detection provides a direct probe of the delay time between star formation and supernova explosion. 
Our estimate of the event rate implies a minimum delay time shorter than 1 Gyr. We also discuss the implications of transient contamination for searches of galaxies in the early Universe.
\end{abstract}

\keywords{\uat{Supernovae}{1668} --- \uat{Type Ia supernovae}{1728} --- \uat{Cosmology}{343} --- \uat{High-redshift galaxies}{734}}


\section{Introduction} 

The James Webb Space Telescope (JWST) has opened a new window into the early Universe through its unprecedentedly deep observations. Among these, one of the most remarkable discoveries enabled by JWST is the numerous discoveries of luminous galaxies in the unexplored high redshift regime up to $z \sim 14$ \citep[e.g.,][]{2024Natur.633..318C,2025ApJ...988...19S,2026OJAp....956033N}. These discoveries have suggested a possible overdensity of luminous galaxies in the early Universe compared with pre-JWST expectations \citep[e.g.,][]{Mason2015,Tacchella2018a,Yung2018}.

In addition, JWST has been expanding the observational frontier of explosive transient phenomena in the Universe, such as supernovae (SNe), up to the redshift of $z \sim 5$ (e.g., JADES survey; \citealt{2025ApJ...990...31D,2024ApJ...972L..13S,2025arXiv250105513C,2026arXiv260712018D,2026arXiv260712028V}). 
For spectroscopically confirmed core-collapse SNe (CC SNe), the current redshift record is a Type II SN (SN II) at $z=5.13$ \citep{2026arXiv260104156C} identified through gravitational lensing. For spectroscopically confirmed Type Ia SNe (SN Ia), the current redshift frontier is $z=2.9$ \citep{2024ApJ...971L..32P} with no strong evidence for redshift evolution in their observed properties. Extending SN Ia observations to even higher redshift is crucial for testing the validity of SN Ia standardization and for establishing SNe Ia as reliable tools for precision cosmology \citep{2025ApJ...981L...9P}. Also, estimating the SN Ia rate in such high-redshift can be a direct probe of the delay time distribution (DTD) between the progenitor formation and its terminal explosion \citep[e.g.,][]{2014AJ....148...13R, 2024A&A...689A.203P}. 

In the high-redshift galaxy searches, candidates are generally selected using multi-band photometric criteria, in particular those designed to identify the distinctive Lyman-limit or Ly$\alpha$ breaks, often referred to as the ``dropout" technique \citep{1995AJ....110.2519S}.
These photometric selection provides a powerful method for identifying high-redshift galaxies. However, even when the imaging data in different filters are obtained at separated epochs, the resulting photometry is typically combined and interpreted as a single, non-variable spectral energy distribution (SED).
As a result, such photometric selections may include transient phenomena such as SNe, as contaminants among high-redshift galaxy candidates. For example, zD0, identified in the Hubble Ultra Deep Field (HUDF), mimicked the SED of a $z>9$ galaxy, but was reported to be a likely transient based on its variability across multiple epochs \citep{2010MNRAS.409..855B}. Such a case for JWST high-redshift galaxy searches was also discussed by e.g., \cite{2023ApJ...947L...1Y, 2025ApJ...979..250D}. 

In this paper, we report the discovery of a SN Ia candidate, \id~at $z\sim4.3$. This object was initially identified as a $z\sim 14$ galaxy candidate \citep{2026arXiv260102861Z}. However, multi-epoch data revealed variability of the source on a timescale of years. The discovery of this object and the observational data are described in Section~\ref{sec:data}. The classification as a transient source, together with the estimates of the SN type, redshift, and phase, is described in Section~\ref{sec:class}. Implications for the SN Ia rate and DTD, as well as the impact on high-redshift galaxy searches, are discussed in Section~\ref{sec:discussion}. Finally, the summary is given in Section~\ref{sec:conclusions}.

All magnitudes used in this paper are given in AB system. We adopt the following cosmological parameters of $\Omega_M$ = 0.3, $\Omega_{\Lambda}$ = 0.7, $H_0$ = 70 ${\rm km \ s^{-1} Mpc^{-1}}$.


\begin{figure*}[]
\centering
	\includegraphics[width=0.32\textwidth]{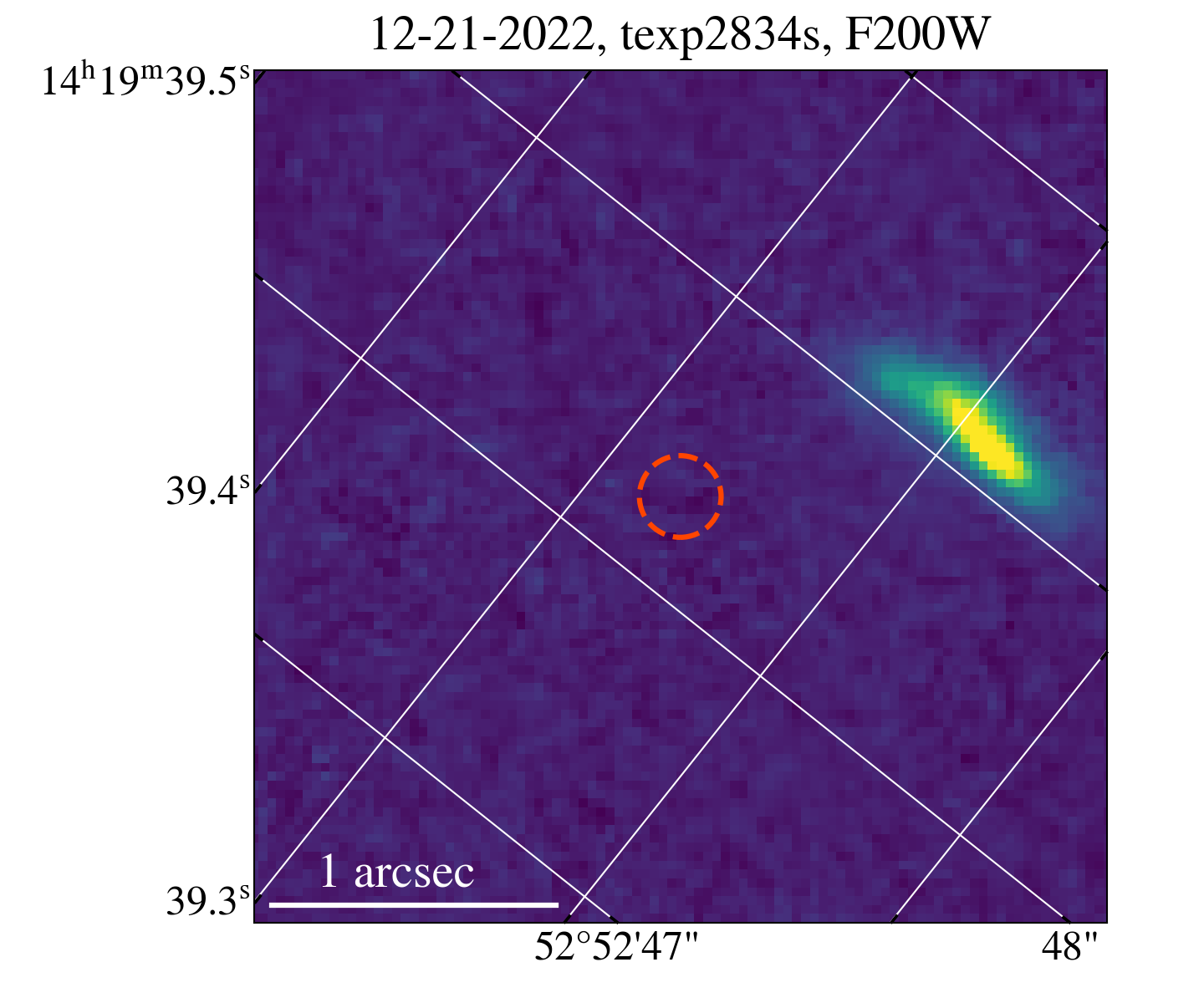}
	\includegraphics[width=0.32\textwidth]{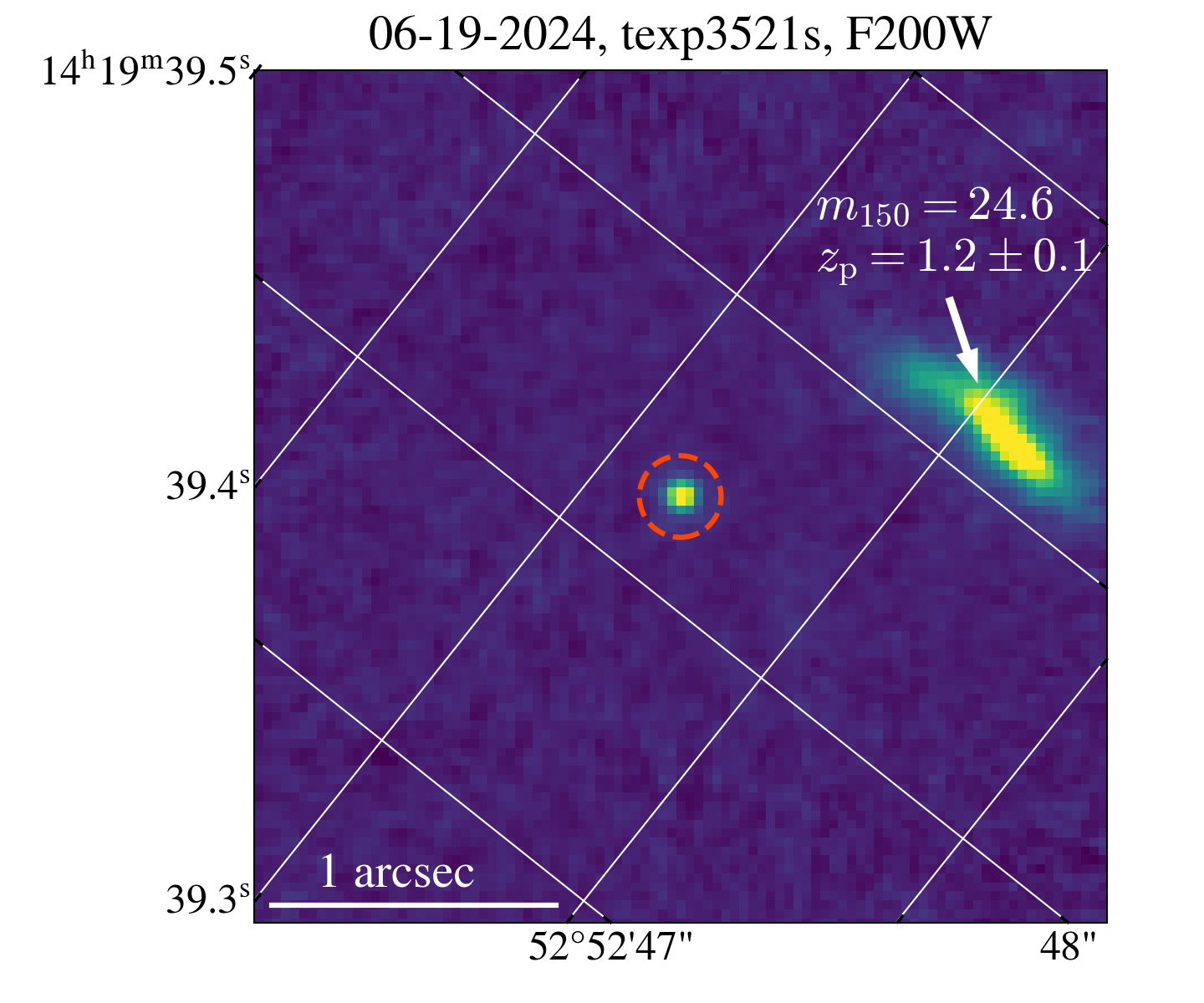}
	\includegraphics[width=0.32\textwidth]{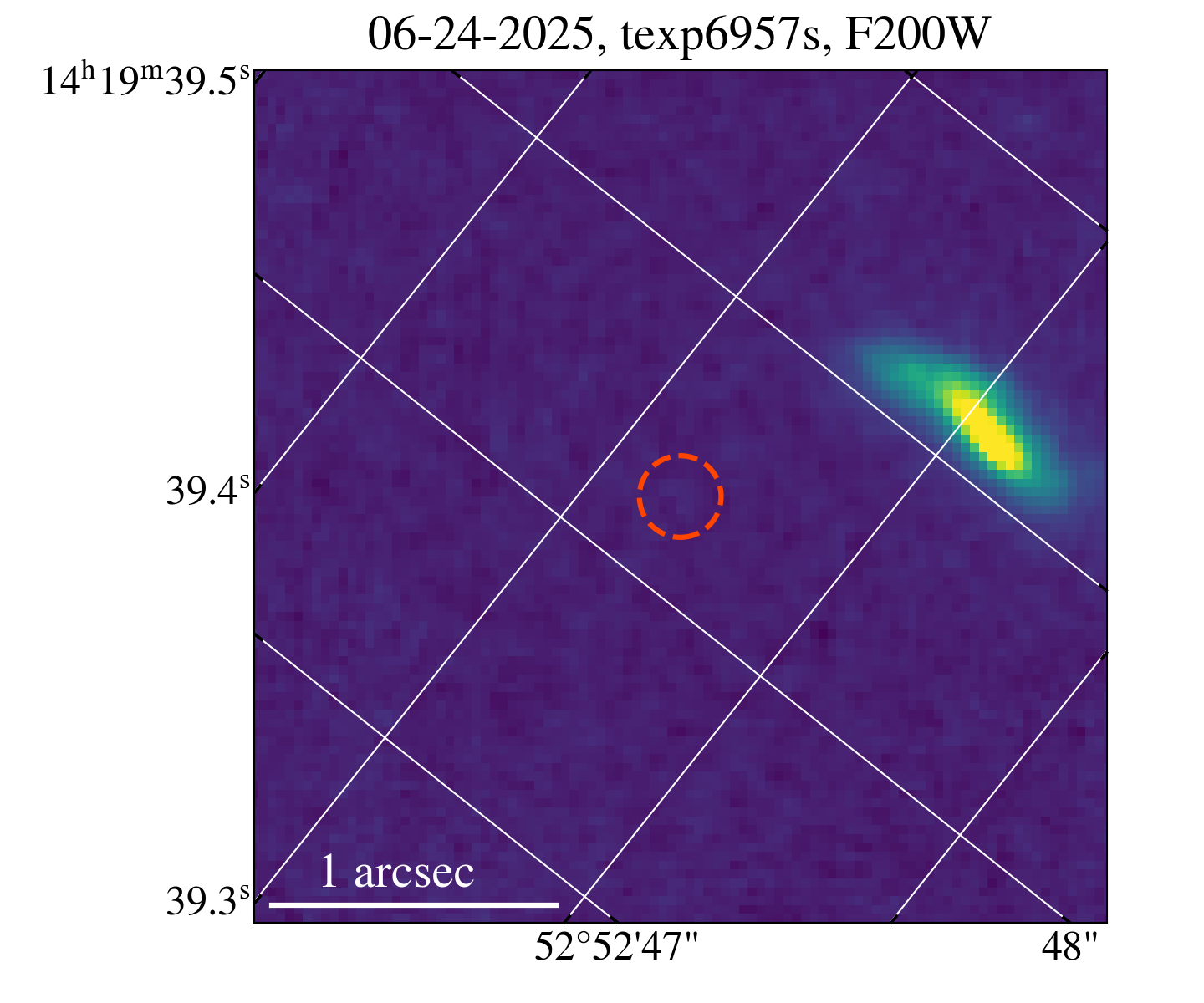}
    \includegraphics[width=\textwidth]{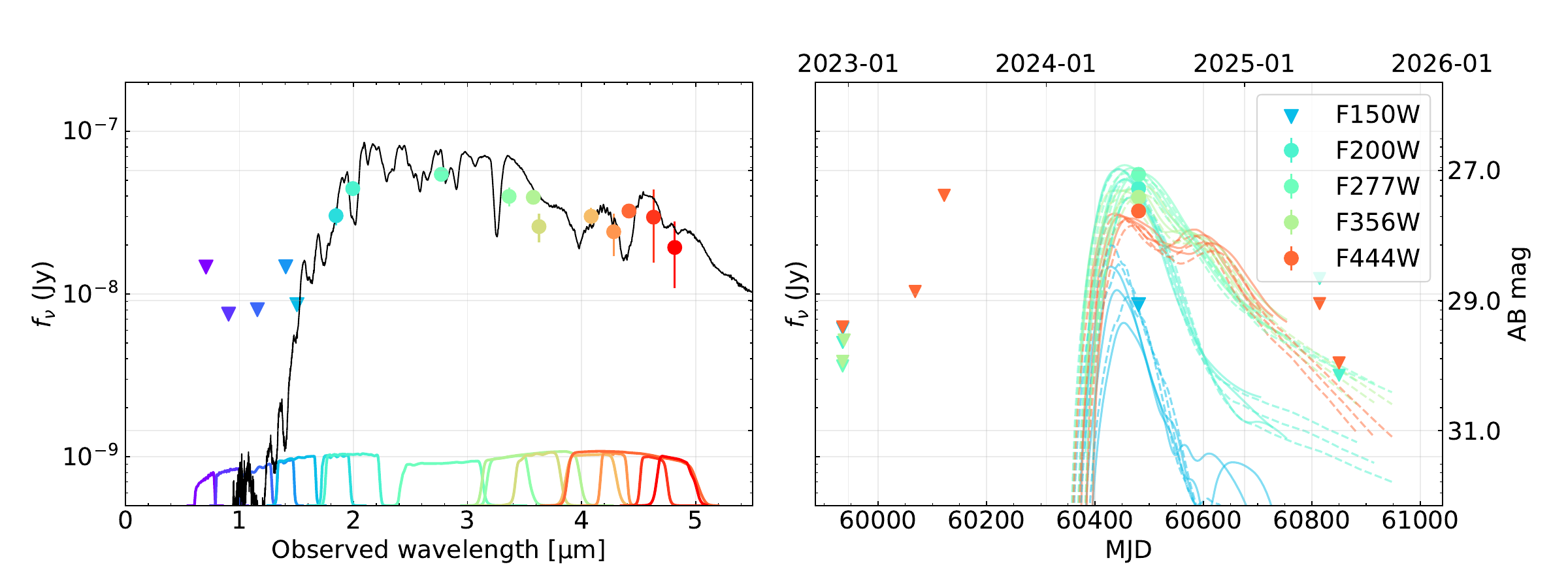}
	\caption{(Upper) JWST NIRCam F200W postage stamps at the position of \id\ (red circles; $r=0.\!''16$) at three different epochs, Dec 2022 (upper left; PID 1345), June 2024 (upper middle; 3990), and June 2025 (upper right; 6434). The image orientation is matched to the June 2024 image. A bright galaxy at $z_{\rm p} = 1.2\pm0.1$ is also seen, at a projected distance of $\sim 8$ kpc from \id. (Lower left) Observed SED of \id. Flux errors for circles correspond to $1\sigma$ and downward triangles correspond to $3\sigma$ upper limits. The spectrum of SN 2011fe at $z=4.3$ (at +3.4 days from the maximum, \citealt{mazzai14}) is also shown for comparison. (Lower right) Light curves of \id\ in the observer frame. For clarity, we show only the wide filters. We also show SN Ia simulated light curves based on {\tt salt2-extended} model and {\tt hsiao} model as solid lines and dashed lines, respectively (see Section \ref{sec:classification} for more details).}
\label{fig:image}
\end{figure*}

\section{The source of interest and data}\label{sec:data}

\id\ was discovered in the CEERS field \citep{2023ApJ...946L..13F} in the Extended Groth Strip (EGS) in JWST/NIRCam imaging obtained on June 19, 2024, or MJD=60480.18, as part of the BEACON Cycle~2 pure-parallel imaging program \citep[PID 3990,][]{2025ApJ...983..152M,2026arXiv260417963K}. 
Based on the full set of 16 NIRCam imaging bands spanning 0.7--4.8\,$\mu$m, \citet{2026arXiv260102861Z} originally identified \id\ as a luminous galaxy candidate at $z_{\rm p}\sim13.7$. 
The source was characterized as a UV luminous ($M_{\rm UV}=-21$\,mag) point-source. 
Two additional NIRCam imaging epochs were obtained at the same pointing approximately one year before and one year after the 2024 observations. 
Neither epoch shows a significant detection at the source position (Figure \ref{fig:image}), despite being sufficiently deep to recover a source with an absolute magnitude of $M_{\rm UV}=-21$\,mag. 
The absence of the same source in those imaging suggests a transient origin, as was briefly mentioned in two independent studies \citep{2026arXiv260111515D,2026arXiv260417963K}.

Our primary analysis focuses on the 16-band NIRCam photometry obtained during the 2024 epoch (hereafter on-epoch), when the source was detected. These observations comprise a total integration time of approximately 8 hours. 
The imaging data from the remaining epochs (off-epochs) are used to constrain the transient light curve in the available filters and to place limits on the underlying host-galaxy emission. 
Those were obtained between December 2022 and June 2026 through programs PIDs~1345 (PI Finkelstein, S., \citealt{2025ApJ...983L...4F}), 2279 (PI Naidu, R., \citealt{2021jwst.prop.2279N}), 2234 (PI Ba\~nados, E., \citealt{2021jwst.prop.2234B}), 6434 (PI Egami, E., \citealt{2025arXiv250503873H}), and 7814 (PI Muzzin, A., \citealt{2025arXiv250719706M}). 
No significant host-galaxy emission is detected in the stacked off-epoch images, and all host-galaxy properties discussed in this work are therefore constrained through upper limits (Appendix~\ref{app:host}). 
A summary of the data used in this study is provided in Table~\ref{tab:photometry} and Table~\ref{tab:data}.

For the 2024 observations, we adopt the publicly available BEACON DR2 photometric catalog \citep{2026arXiv260417963K}, and therefore our photometry is consistent with that reported by \citet[][]{2026arXiv260102861Z}. We summarize the fluxes in Table~\ref{tab:photometry}. Data from all other programs were retrieved from the MAST archive in the form of Level-3 products. Limiting magnitudes at the position of \id\ were measured using a circular aperture with a radius of ($r=0.\!''16$) and reported in Table~\ref{tab:data}.

We note that \citet{2026arXiv260102861Z} investigated the possibility that \id\ is a foreground object (i.e., a brown dwarf) and found no compelling evidence supporting such an interpretation based on template-fitting analyses. Furthermore, although a nearby foreground source could exhibit measurable proper motion over the time baseline considered here, we detect no significant positional shift during the 2024 observations and find no counterpart at or near the source location in any of the off-epoch imaging. The absence of a persistent source in the pre- and post-event data further disfavors a foreground origin.


\section{Supernova classification} \label{sec:class}

\begin{figure}[]
\centering
	\includegraphics[width=1\columnwidth]{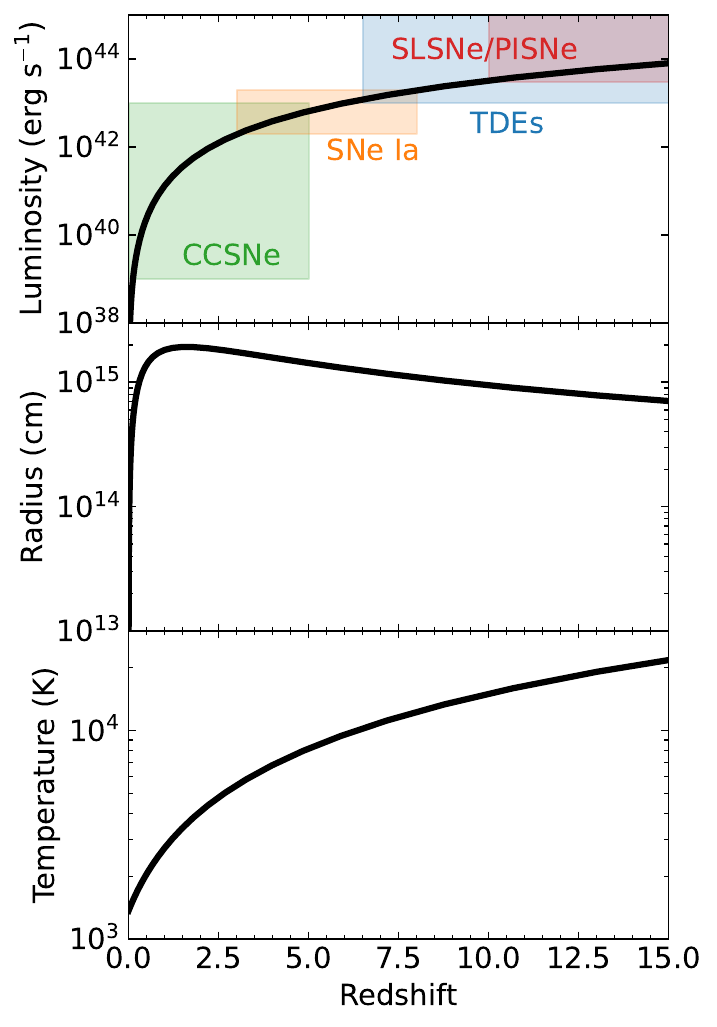}
	\caption{Luminosity (top), radius (middle), and temperature (bottom) derived from the blackbody fitting as a function of redshift.
    In the top panel, the shaded regions show typical luminosity ranges (from the maximum luminosity to $\sim 10\%$ of the maximum) and corresponding redshift ranges for different types of transients. 
    }
\label{fig:param}
\end{figure}

\subsection{General constraints}
\label{subsec:general}

We first broadly constrain the nature of \id\ as a transient object by fitting the observed SED with a single blackbody function.
Figure \ref{fig:param} shows the estimated luminosity, radius, and temperature of the blackbody as a function of assumed redshift. 
As naturally expected, the source should be intrinsically more luminous when a higher redshift is assumed.
Also, to reproduce the observed SED, the temperature tends to be higher at a higher redshift as the SED peak shifts to shorter wavelength in the rest frame.
Estimated radius and temperature are broadly consistent with those expected for SNe.

From the characteristic peak luminosity ranges of various types of transients (top panel of Figure \ref{fig:param}), 
we can obtain first-order estimates of a redshift for different types of transients.
A likely redshift is $z \leq 5$ for CC SNe, while it is $z \simeq 3-8$ for SNe Ia.
For more luminous transients such as tidal disruption events (TDEs), superluminous SNe (SLSNe, including pair-instability SNe), a likely redshift is even further, $z \geq 8$.

First, we can filter out incompatible transient scenarios based on the observed temporal variability of \id, namely, a change of 3 mag over one year in the observer frame, corresponding to a flux variation of about one order of magnitude (Figure \ref{fig:image}). For typical SLSNe and TDEs, owing to their luminosities and the cosmological time-dilation effect, light curves of these populations readily overshoot the off-epoch upper limits (see Figure \ref{fig:lc-model} and Appendix \ref{app:phot}). However, we note that rapid transients such as fast blue optical transients (FBOTs) or rapidly evolving TDEs (e.g., AT 2020wey, \citealt{2023A&A...673A..95C}), cannot be completely ruled out due to their rapidly evolving light curve properties. AGN variability may also provide a possible explanation for the transient nature of the source (e.g., \citealt{2024ApJ...971L..16H}). However, the absence of a persistent source in the off-epoch images disfavors this interpretation in terms of the typical stochastic AGN variability. 

Furthermore, from the consideration of the expected event rate,
an interpretation as luminous transient populations at $z \geq 8$ is very unlikely.
As discussed in Section \ref{sec:rate}, \id\ was detected in the JWST images over 400 arcmin$^2$. Although the entire area does not have uniform multi-epoch coverage, the number of expected sources over the full area provides the most conservative estimate of the detectability of rare transient populations that could account for \id.
Assuming that the CC SN rate follows the cosmic star formation rate, the expected number of total CC SNe at $z \geq 8$\footnote{Although CC SNe at such high redshift would be too faint to be detected even with JWST depths, we refer to their event rates for comparisons.} is 1-10 per 400 arcmin$^2$ per 1 yr in the observer's frame.
The event rates of TDEs, SLSNe, and FBOTs are smaller than the total CC SN rate at least by 3 orders of magnitudes (TDEs, \citealt{2023ApJ...955L...6Y}; SLSNe, \citealt{cooke2012, moriya2019}; FBOTs, \citealt{ho2023}). 
This indicates that the expected number of these events in our search area is $10^{-3} - 10^{-2}$ in 1 year in the observer's frame.
Therefore, in the following section,
we perform photometric classification for more common SNe, i.e., CC SNe and SNe Ia.

\begin{figure*}[]
\centering
\includegraphics[width=0.95\textwidth] {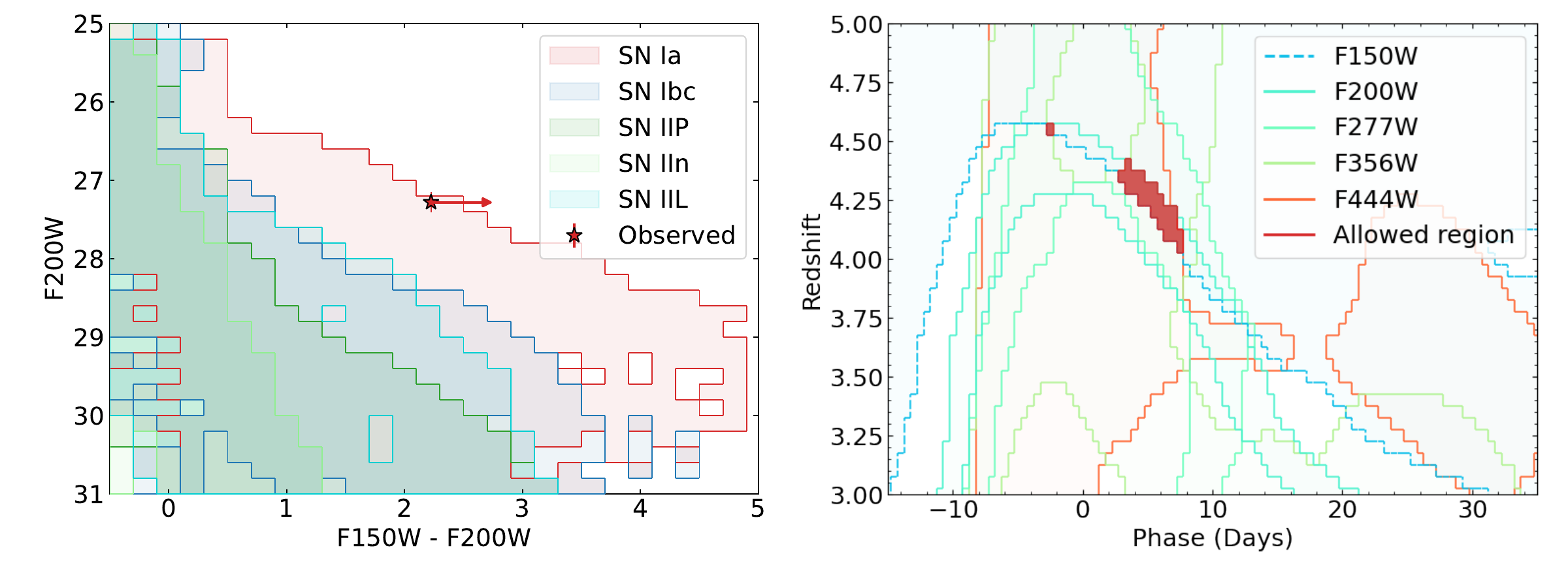}
	\caption{(Left) Color-magnitude diagram of simulated transients.  \id~ is shown as a red star. The shaded regions indicate the color-magnitude space covered by each SN type based on the simulated light curves. (Right) The allowed phase and redshift region obtained by comparing simulated SN Ia light curves with the observed photometry of \id.}
\label{fig:color-mag}
\end{figure*}

\subsection{Photometric classification}
\label{sec:classification}

We here constrain transient types and redshifts of the source using the available photometric information. We simulate light curves for SNe Ia and CC SNe (Ibc, IIP, IIL, and IIn) using {\tt SNCosmo}, a Python library for SN cosmology \citep{sncosmo}. For SN Ia, we use the built-in model, {\tt salt2-extended} \citep{2018PASP..130k4504P} since the model has wide wavelength (1700~\AA\ to 24990~\AA) and phase (rest-frame days relative to the light curve peak,\ $-20$ days to $+50$ days) coverage. The light curve models are expressed with the stretch parameter $x_{1}$ and color parameter $c$. The peak absolute magnitude and these parameters are related by $M_{B} = m_{B} - \mu - \alpha x_{1}  - \beta c$. We used $\alpha = 0.141$, $\beta = -3.101$, $x_1 = 0.945$, $c=-0.043$, and $m_{B}-\mu = -19.05$ following \citet{2014A&A...568A..22B} and \citet{ 2016ApJ...822L..35S}. For CC SNe, we adopt built-in sources {\tt nugent-sn1bc} \citep{2005ApJ...624..880L}, {\tt nugent-sn2p}, {\tt nugent-sn2l}, and {\tt nugent-sn2n}~\citep{1999ApJ...521...30G}. The wavelength and phase coverages are listed in Table~\ref{tab:template_coverage}. 

We simulate light curves over a redshift range of $3 \lesssim z\lesssim 8$ for SNe Ia and $0 \lesssim z \lesssim 5$ for CC SNe based on the discussion in Section~\ref{subsec:general}. Since the survey area is small and SNe with typical luminosities are expected to dominate, we do not vary the peak magnitude of each SN class, but instead assign a representative peak absolute magnitude to each type (Table~\ref{tab:template_coverage}).

Since built-in models cannot cover the entire wavelength observed by JWST (longer than $2.5~\rm \mu m$, for SNe in the relatively low-redshift regime, $z\lesssim1$), we first focus on light curves at wavelengths shorter than $2.5~\rm \mu m$. In particular, we focus on the color-magnitude diagram F150W-F200W vs F200W (left panel of Figure \ref{fig:color-mag}).
For SNe Ia at $z\sim3-5$, F150W-F200W color is sensitive to the absorption features produced by Fe-group elements in the rest-frame UV wavelength range ($\sim$3000-4000~\AA).  
By comparing the distribution of the SN templates in the color-magnitude diagram with \id, we can largely eliminate CC SNe scenarios. This is because CC SNe generally show weaker Fe-group absorption features. Although similar spectral breaks between F150W and F200W could arise from a cooler blackbody continuum at later phases, the luminosity in such phases would be too faint to reproduce the observed brightness.
Instead, we confirm that SNe Ia are broadly consistent with \id. 

We then focus on the SN Ia interpretation and constrain phase and redshift. We investigate the combinations of phase and redshift that can explain the photometry of \id. Here to obtain the broad constraints, we explore the parameter space of phase and redshift to reproduce the observed photometry within $\pm3\sigma$ for wide filters.

The right panel of Figure~\ref{fig:color-mag} shows the allowed region in the parameter space of the SN Ia phase and redshift. The result suggests that SN Ia around peak time ($0 \lesssim t_{0} \lesssim 10$ days) at redshift $4\lesssim z\lesssim 4.5$, can reproduce the photometry of \id~well. 
Notably, in addition to the color, the observed brightness is well matched by a normal SN Ia with a typical absolute magnitude at this redshift.

Moreover, the SN Ia models at these redshifts and phases also naturally reproduce the observed light curve evolution. In the lower right panel of Figure \ref{fig:image}, the simulated light curves of the SN Ia models corresponding to the selected redshifts and phases are shown with the light curves of \id. We add the simulated light curves with the {\tt hsiao} models \citep{2007ApJ...663.1187H}, since the {\tt salt2-extended} model does not cover a sufficiently wide time range to be compared with the data points at $\Delta t\gtrsim 300~\rm{days}$ in the observer's frame.

We note that there is a galaxy at photometric redshift of $z_{\rm p} = 1.2\pm0.1$ located at a projected distance of $\sim 8$ kpc from \id\ (Figure~\ref{fig:image}). If we assume that this is instead the host galaxy of \id, CC SNe at $z\sim1.2$ can be the nature of this object based on the discussion on luminosity in Section 3.1. However, typical CC SNe at $z\sim1$ cannot reproduce the color of \id\ as shown in the left panel of Figure \ref{fig:color-mag}.
This $z_{\rm p} \approx 1.2$ galaxy may potentially gravitationally lens and magnify the background source. However, we do not expect this to have a large impact as the foreground source is faint and low mass ($\log M/M_\odot =8.43$). Following the approach of \citet{Mason2015a} we estimate the magnification is $<+15\,\%$.

\begin{figure*}[]
\centering
	\includegraphics[width=0.48\textwidth]{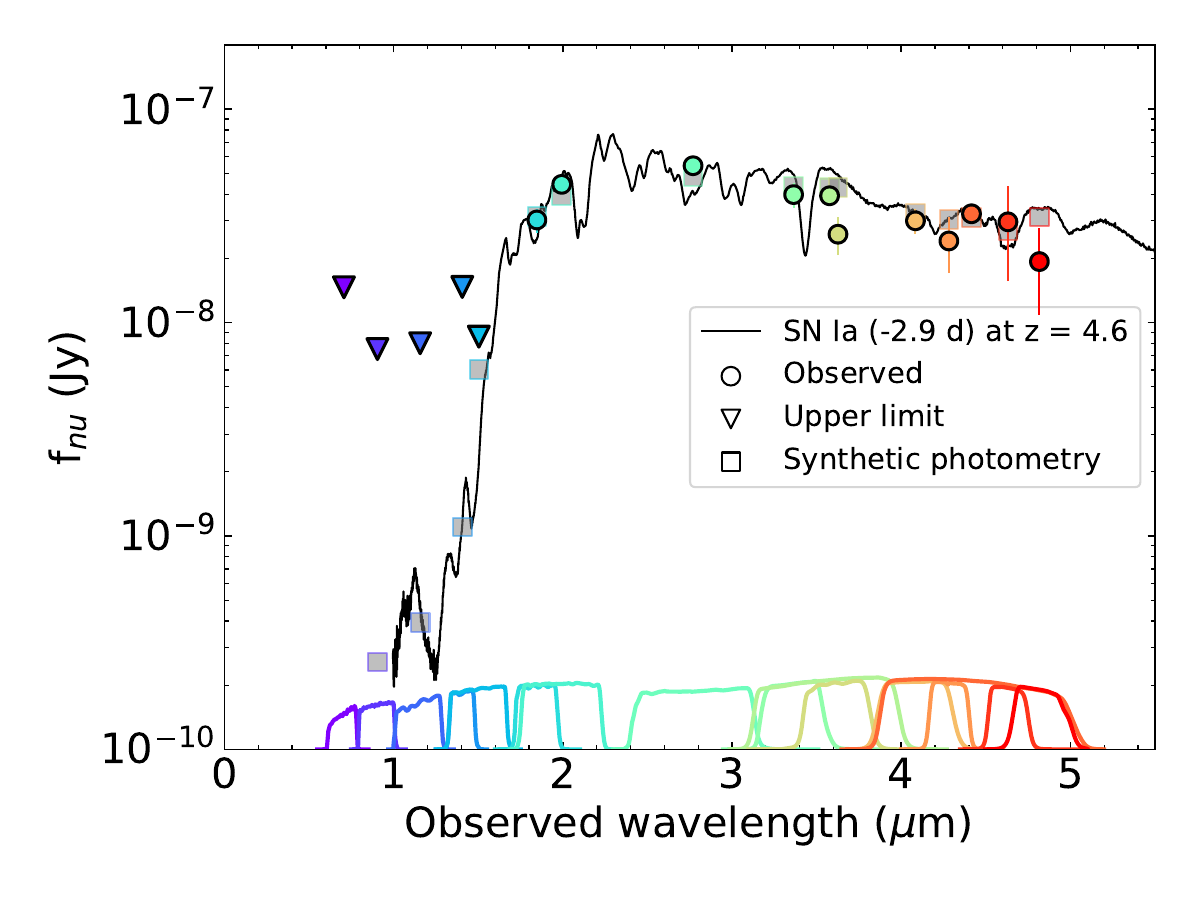}
    \hfill
    \includegraphics[width=0.48\textwidth]{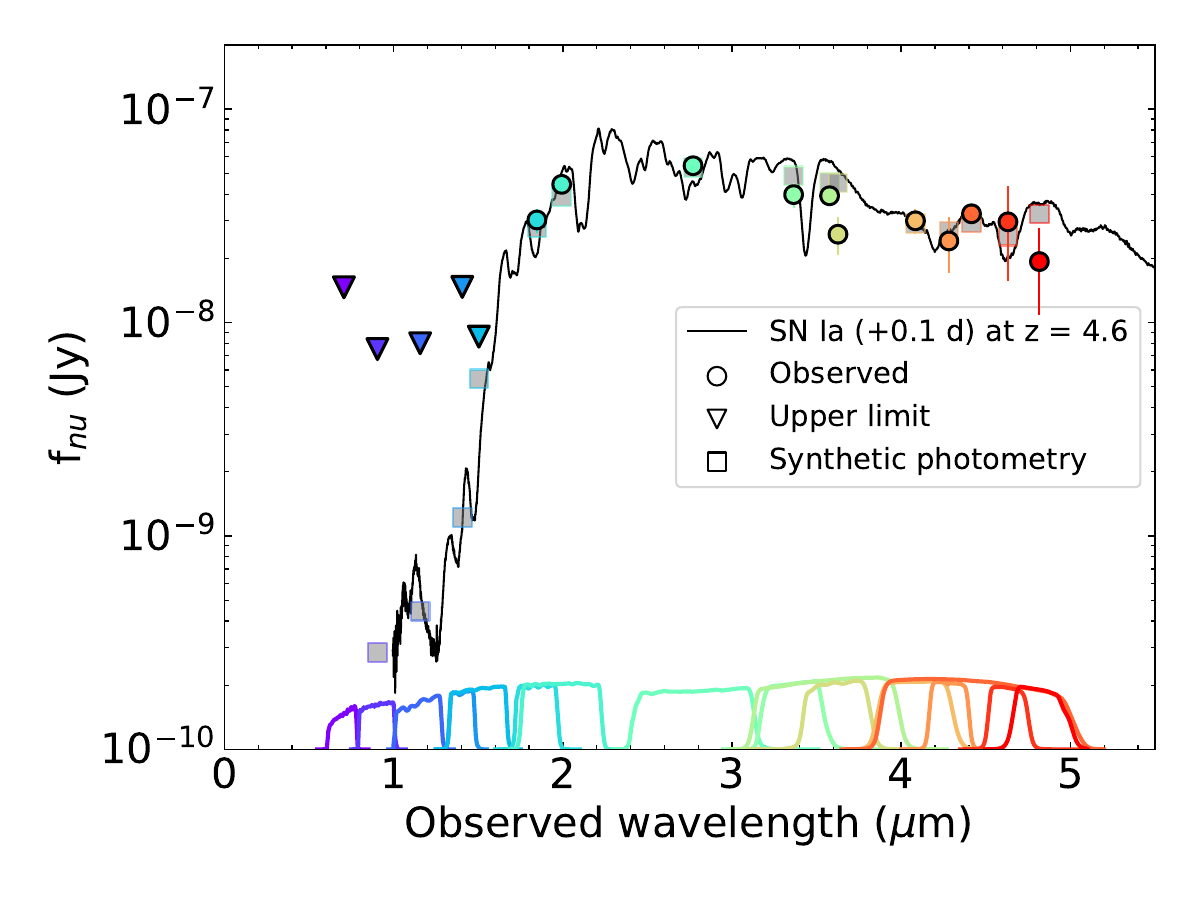}
    
    \includegraphics[width=0.48\textwidth]{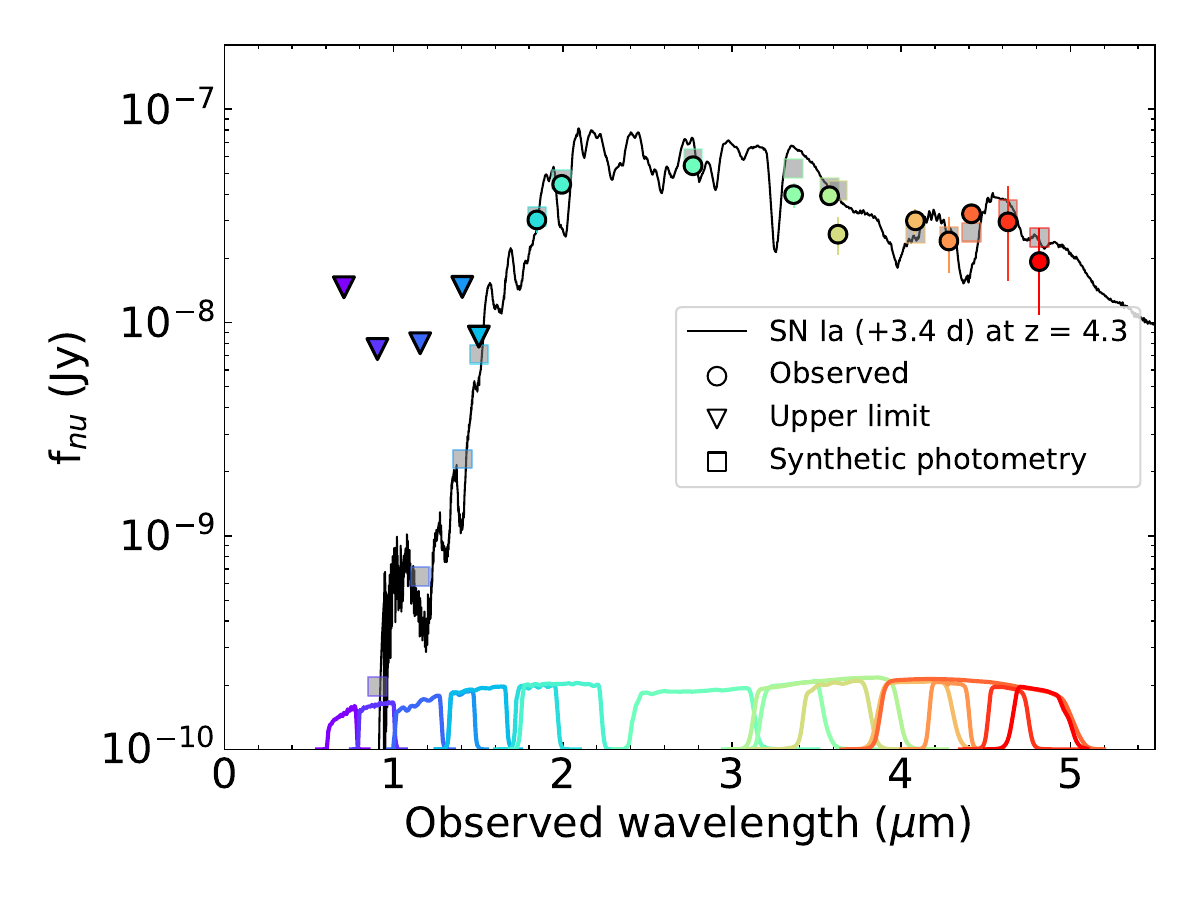}
    \hfill
    \includegraphics[width=0.48\textwidth]{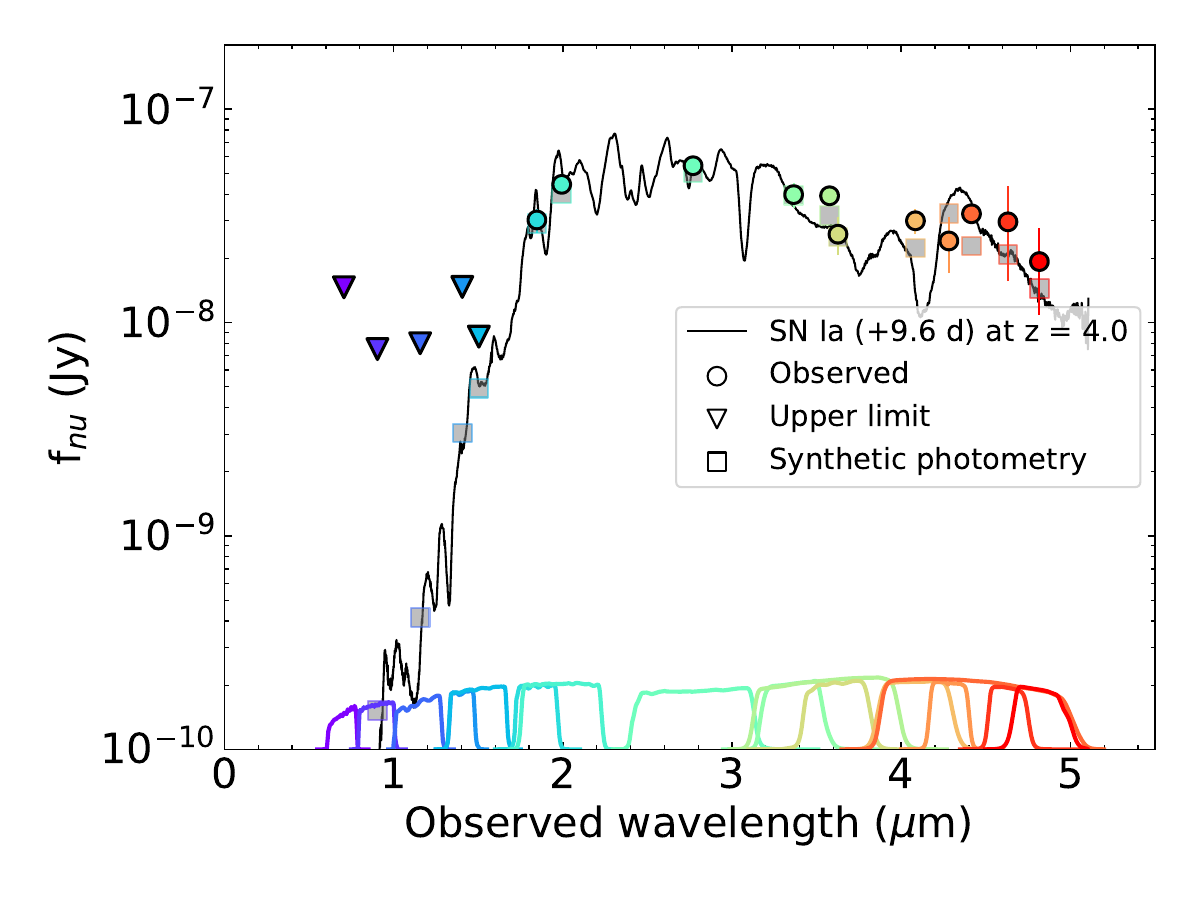}
	\caption{Observed SED (circles and triangles) compared with the spectra of SN 2011fe (black curves, \citealt{mazzai14}).
    Synthetic photometry of the spectrum are shown with squares.
    Each panel shows the spectrum at a different phase at the best-fit redshift:
    spectrum at $-2.9$ days from the maximum ($z=4.6$, upper left panel), 
    spectrum 
     at $+0.1$ days ($z=4.6$, upper right panel), 
    spectrum at +3.4 days ($z=4.3$, lower left panel), and spectrum at $+$9.6 days ($z=4.0$, lower right panel). 
    }
\label{fig:sed_best}
\end{figure*}

\subsection{SED fitting}
\label{sec:sed}

We further constrain the redshift of \id\ by using the multi-band photometry and SN Ia spectral templates.
We use spectra of the prototypical SN Ia 2011fe taken with the Hubble Space Telescope at $-$2.9, +0.1, $+$3.4, and $+9.6$ days from the maximum, covering from the rest-frame UV to NIR \citep{mazzai14}.
These epochs are selected based on the allowed parameter space as discussed in Section \ref{sec:classification}.
The JWST photometry is compared with the synthetic photometry from these template spectra by varying the redshift.
We do not apply any correction to the intrinsic luminosity.

A SN Ia spectrum at $+$3.4 days at $z = 4.3$ (motivated by the photometric classification) reproduces the observed SED remarkably well.
The left bottom panel of Figure \ref{fig:sed_best} shows the best fit result, giving $\chi^2$/dof = 3.0.
The observed flux drop at the shorter wavelength
is reproduced by the strong absorption by Fe-group elements in the rest-frame UV wavelength
(this relatively sharp drop mimics Lyman break for high-redshift galaxy searches).
The slope at the longer wavelength also nicely
agrees with the pseudo continuum of SN Ia spectrum.
In fact, the relatively steep rest-frame UV absorption gives the strong constraints on the redshift.
At lower/higher redshift, the absorption features appear at shorter/longer wavelengths in the observer's frame, which does not agree with the observed sharp cutoff.

The best-fit redshift depends on the phase of the adopted spectra since the wavelength of the rest-frame UV break slowly evolves with time (Figure \ref{fig:sed_best}, see also Appendix \ref{app:sed}).
The spectra at $-$2.9 and $+0.1$ days give a somewhat higher redshift estimate ($z \simeq 4.6$), while the spectrum at $+9.6$ days gives a smaller redshift estimate ($z \simeq 4.0$).
Thus, by including the uncertainty in the epoch, the redshift of \id\ is likely to be $z \simeq 4.0-4.6$.
In Appendix \ref{app:host}, we discuss properties of the host galaxy at this redshift based on the non-detection at off-epoch.




\section{Discussion}
\label{sec:discussion}

\subsection{Event rate and Delay Time Distribution}
\label{sec:rate}

SN Ia rate as a function of redshift provides us with an important probe to constrain their progenitors as the event rate depends on the DTD. 
Here, we estimate the event rate of SNe Ia at redshift $z\sim 4.3$. 
Using the detection efficiency $\epsilon$, comoving volume $\Delta V$, and survey duration~$\Delta T$, the event rate $r$ can be estimated as:
\begin{align}
r = \frac{1}{\epsilon\Delta V(z) \Delta T(z) }.
\end{align}

The SN Ia candidate \id\ was discovered 
as part of the BEACON program, which spans $\sim 400~{\rm arcmin^2}$ \citep{2025ApJ...983..152M, 2026arXiv260417963K}. 
Although multi-epoch data are available only for a fraction of the survey fields, we adopt the total survey area for our rate estimation. For those without multi-epoch data, we are not allowed to identify a transient event. In addition, even for the fields with multi-epoch imaging, we may miss transients if they are superposed on bright extended galaxies, whereas \id~ was observed as an isolated source. These effects may be included in the detection efficiency $\epsilon$. Nevertheless, to be conservative, we adopt $\epsilon = 1$ in this study, which gives the lowest event rate.

To estimate the survey volume, we consider the redshift range and phases over which SN Ia can be detected. 
The high-redshift ($ 13\lesssim z < 18$) galaxy selection corresponds to the criteria on the signal to noise ratios (SNR) for each band such as ${\rm SNR_{F277W}} > 4$, ${\rm SNR_{F150W}} < 2$, ${\rm SNR_{F115W}} < 2$, and  ${\rm SNR_{F090W}} < 2$. Since the imaging depth varies across the BEACON field, we adopt the median limiting magnitude in each filter, 27.23, 27.82, 27.97, and 28.62 mag for $5\sigma$ in F090W, F115W, F150W, and F277W, respectively, to estimate the typical survey volume and duration. We apply the same cut for SN Ia simulated light curves and estimate the redshift and phase coverage as shown in Figure \ref{fig:beacon-ia}.
From the estimated coverages, we find $\Delta V\sim 6.2\times10^{6}~{\rm Mpc^3}$ and $\Delta T \sim 40~\rm{days}$ for SNe Ia in rest frame.
By using these values, we estimate the event rate of SN Ia at $z\sim 4.3$ as  $1.7^{+3.9}_{-1.4} \times 10^{-6}~\mathrm{Mpc^{-3}\,yr^{-1}}$ with $1\sigma$ confidence intervals \citep{1986ApJ...303..336G}. 
As discussed above, this should be a lower limit of the actual SN Ia rate at this redshift.  

Next, we compare this result with the expected SN Ia rate expected from the DTD of SN Ia \citep{2014ARA&A..52..107M}. Since the cosmic age at $z\sim4$ is about 1.5 Gyr, the SN Ia rate at this redshift is very sensitive to the minimum delay time.
We adopt the cosmic star-formation rate density model presented by \citet{2017ApJ...840...39M}: 
\begin{equation}
\psi(z)=0.01\,{(1+z)^{2.6}\over 1+[(1+z)/3.2]^{6.2}}\,\sfrd.
\label{eq:sfrd}
\end{equation}
We assume the form for the DTD as follows \citep{2011NatCo...2..350H, 2017ApJ...848...25M}:
\begin{equation} 
\Psi_{\rm Ia}(t) = 
    \begin{cases} 
    0 & (t < t_{\rm min}), \\[4pt] 
    A \left( \dfrac{t}{t_{\rm min}} \right)^{-1} & (t_{\rm min} \leq t). 
    \end{cases} 
\end{equation}
The normalization factor $A$ determined by requiring that the DTD integrated over a Hubble time reproduces the SN Ia production efficiency $N/M_{*}$. We adopt $N/M_{*} =1.3\times10^{-3}$, as estimated by \citet{2012MNRAS.426.3282M}. Then, we obtain the SN Ia rate evolution model by convolving the star formation history and the DTD as follows:
\begin{equation}
R_{\rm Ia}(z)= \int_{t_{\rm min}}^{t_{\rm age} (z)}\psi(t(z) - \tau)\Psi_{\rm Ia}(\tau)d\tau.
\label{eq:ria}
\end{equation}
We compare the estimated event rate with the rate models adopting different minimum delay times $t_{\rm min}=0.1,~0.3,~0.5,$ and$~1.0~{\rm Gyr}$ in DTD. The estimated event rate exceeds the model predictions for $t_{\rm min} = 1.0~{\rm Gyr}$, suggesting a minimum delay time of $t_{\rm min} < 1~{\rm Gyr}$. 
Note that our estimated event rate is a lower limit, and the true rate would be even higher, preferring a shorter delay time.
Our results demonstrate that the observation of even a single SN Ia at $z\sim4.3$ already provides an interesting constraint on the minimum delay time.

\begin{figure}[]
\centering
	\includegraphics[width=1\columnwidth]{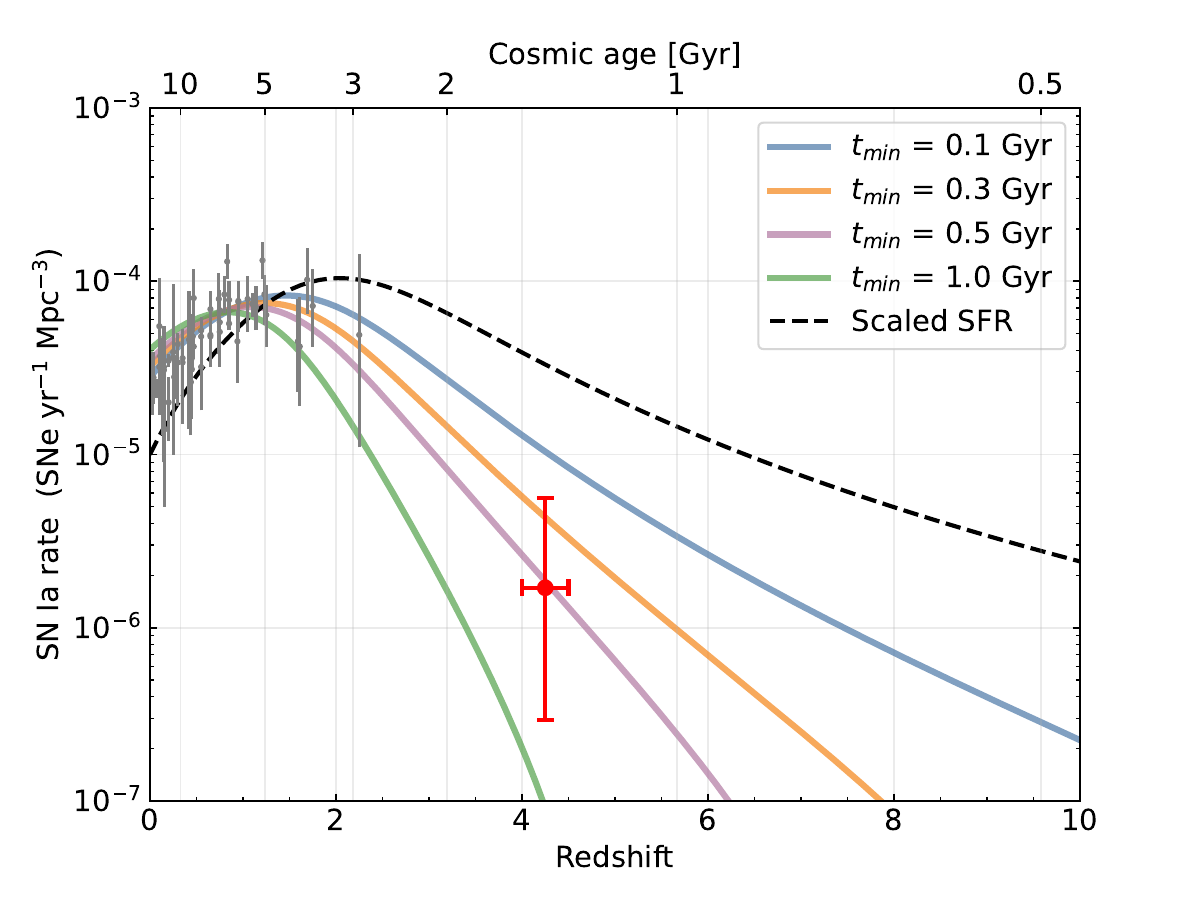}
	\caption{Event rate from \id\ (red circle) compared with SN Ia rate predictions with different minimum delay times. The error bar corresponds to the $1\sigma$ confidence interval.
    }
\label{fig:rate}
\end{figure}

\subsection{Implication for high-redshift galaxy searches}
\label{sec:highz}
Given the currently limited survey volume of JWST, the misidentification of even a single SN as a high-redshift galaxy can have a significant impact on estimates of the galaxy population. 
Owing to the small number of galaxies currently known at comparable redshift and luminosity ($M_{\rm UV}\sim -22$\,mag), the inclusion of this single source would have a non-negligible effect on inferred number densities. Using the small-number Poisson statistics of \citet{1986ApJ...303..336G}, we estimate that its inclusion would increase the measured galaxy number density by $\sim0.25$\,dex (a factor of $\sim1.8$) when the upper bound is compared.

In high-redshift galaxy searches, contamination from low-redshift dusty galaxies and Galactic cool stars has long been recognized and extensively studied \citep[e.g.,][]{2005ApJ...622..319P,2020ApJ...904...50M,2025arXiv251000111H,2026arXiv260423668B,2026arXiv260702354I}. In contrast, contamination from transient sources such as SNe has received relatively little attention until recently \citep{2023ApJ...947L...1Y,2025ApJ...990...31D}. Unlike traditional contaminants, SNe can temporarily exhibit colors that closely mimic the Lyman-break signature used to identify high-redshift galaxies, making them difficult to distinguish in single-epoch observations. 
One notable example is {\it Capotauro}, a candidate $z\sim30$ galaxy \citep{2026A&A...706A.364G}. However, \citet{2026OJAp....962107F} later present evidence that {\it Capotauro} may instead be a PISN at $z\sim15$ \citep[see also][]{2026ApJ..1001....3J}. One obvious yet crucial lesson from these studies, and ours too, is an asymmetric contamination effect i.e. the far more common lower-redshift SNe can migrate into higher-redshift galaxy samples, where even a small number of contaminants can have a catastrophic impact on inferred number densities.

Although a number of JWST deep fields have now been observed at multiple epochs, allowing variability tests for sources originally selected from single-epoch imaging, the total area covered by such data remains limited. Most JWST extragalactic surveys still rely on single-epoch observations over substantially larger areas, leaving them vulnerable to contamination from transient sources. Repeat imaging therefore remains the most robust method for identifying and removing such contaminants from high-redshift galaxy samples.


\section{Conclusions} 
\label{sec:conclusions}

We report the discovery of a SN Ia candidate at $z\sim4.3$ with JWST, which was initially identified as a high-redshift galaxy candidate at $z\sim 14$. By comparing the colors and magnitudes of \id\ with simulated light curves of transients, we classify \id\ as a SN Ia candidate. The SED fitting result also suggests that the object is a typical SN Ia at $z\sim4.3$ around peak brightness. The rest-frame UV strong absorption by Fe-group elements seen in SN Ia spectra at $z\sim4$ can mimic Lyman break in galaxies at $z\sim14$. We estimate the event rate of SN Ia around $z\sim4.3$ as $1.7^{+3.9}_{-1.4} \times 10^{-6}~\mathrm{Mpc^{-3}\,yr^{-1}}$. 
The estimated SN Ia rate prefers the minimum delay time $t_{\rm min} < 1~{\rm Gyr}$. 
Our results demonstrate that observations of even a single SN Ia at this redshift can provide an interesting constraint on the DTD.

The identification of this SN Ia candidate among high-redshift galaxy candidates also has implications for high-redshift galaxy searches. The inclusion of transients among high-redshift galaxy candidates would increase the measured galaxy number density by $\sim 0.25~{\rm dex}$. Multi-epoch imaging is crucial for removing transient contaminants from high-redshift galaxy searches. 

\section*{Acknowledgements}
Some of the data presented in this paper were obtained from the Mikulski Archive for Space Telescopes (MAST) at the Space Telescope Science Institute. We acknowledge the teams of the JWST observation programs IDs~1345, 2279, 2234, 3990, 6434, and 7814 for their dedicated work in designing and planning these programs. We thank Iair Arcavi, Hirofumi Noda and the JWST-BEACON collaboration for useful discussion. 

This research was partially supported by JSPS KAKENHI Grant Nos. 25KJ0556 (ST), 24K00668 (KK). S.T. acknowledges support from the Graduate Program on Physics for the Universe (GP-PU) at Tohoku University. AJB acknowledges funding from the “FirstGalaxies” Advanced Grant from the European Research Council (ERC) under the European Union’s Horizon 2020 research and innovation program (Grant agreement No. 789056).

\input{flux}
\input{limmag}

\appendix

\section{Light curve comparison}
\label{app:phot}

\begin{figure}[]
\centering
\includegraphics[width=0.9\columnwidth] {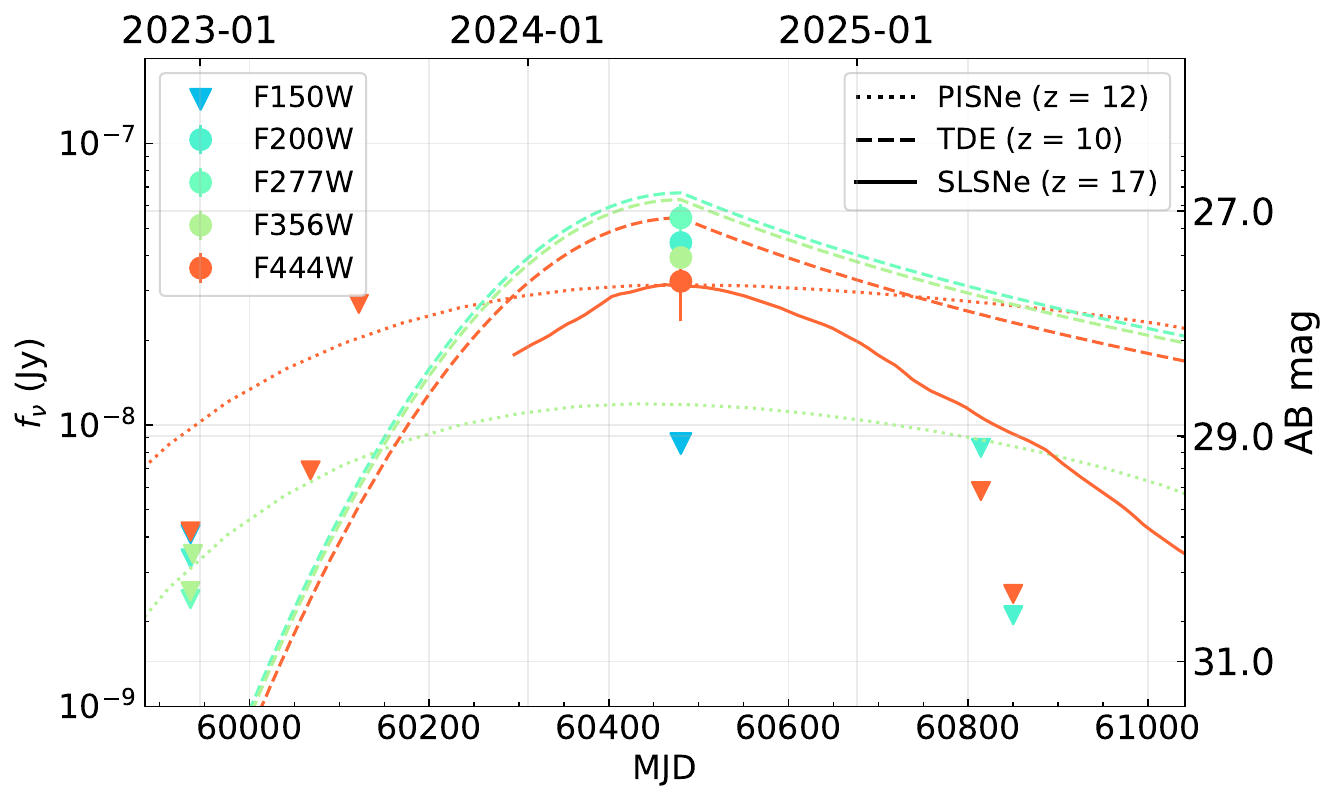}
	\caption{Light curves of \id\ compared with simulated light curves of luminous and rare transient populations. The solid, dashed, and dotted lines show SLSNe, TDEs, and PISNe, respectively.
    }
\label{fig:lc-model}
\end{figure}

We compare the observed light curves of \id\ with simulated light curves of SLSNe, PISNe, and TDEs. We adopt the mean SLSNe model from \citealt{2024MNRAS.535..471G} which spans the phase range of $-23$ and 243 days and wavelength range of $2,100$ and $16,000$~\AA. For PISNe, we adopt {\tt He110} as a representative model from Helium-core PISN light curves calculated by Sasha Kozyreva\footnote{https://wwwmpa.mpa-garching.mpg.de/ccsnarchive/data/Kozyreva/PISN/}. We use the simple model from \citet{2019ApJ...878...82V} with a gaussian rise and a power-law decay for the long-term light curve of TDEs. We adopt $L_{\rm peak}=1.5\times10^{44}~\rm{erg/s}$ and $T_{\rm peak}=20,000~{\rm K}$ with a power-law index $p=-5/3$ as a representative of observed TDE samples \citep{2023ApJ...942....9H}. We set the redshift for each model to reproduce the on-epoch observed magnitude with the peak of the light curve. Figure \ref{fig:lc-model} shows the comparison between \id\ and the light curve models. The long light curve durations of these populations make it difficult to explain \id\ in addition to the rarity of these populations.

\begin{table*}
\centering
\caption{Summary of the transient models used in {\tt SNCosmo}.}
\label{tab:template_coverage}
\begin{tabular}{llrcccc}
\hline
Type & Source / template & Phase range & Wavelength range & Redshift range & Peak magnitude$^{a}$ & Reference \\
\hline
SN Ia  & \texttt{salt2-extended}        & $-20$ -- $+50$ d & 1700--24999 \AA & $3.0$--$8.0$ & $-19.32$ 
&  \citet{2018PASP..130k4504P} \\
SN Ibc & \texttt{nugent-sn1bc} & $0$ -- $+85$ d & 1000--25000 \AA & $0.1$--$5.0$ & $-17.54$
& \citet{2005ApJ...624..880L} \\
SN IIP & \texttt{nugent-sn2p}  & $0$ -- $+411$ d & 1000--25000 \AA & $0.1$--$5.0$ & $-16.80$ 
&\citet{1999ApJ...521...30G} \\
SN IIL & \texttt{nugent-sn2l}  & $0$ -- $+411$ d & 1000--25000 \AA & $0.1$--$5.0$ & $-17.98$
& \citet{1999ApJ...521...30G} \\
SN IIn & \texttt{nugent-sn2n}  & $0$ -- $+237$ d & 1000--25000 \AA & $0.1$--$5.0$ & $-18.62$
&\citet{1999ApJ...521...30G} \\
\hline
\end{tabular}
\tablecomments{(a) Peak absolute magnitudes in B-band. For SNe Ia, we adopt the parameters in \citet{2014A&A...568A..22B} and \citet{2016ApJ...822L..35S}, while for CC SNe, we use the peak absolute magnitudes from \citet{2014AJ....147..118R}.}
\end{table*}

\section{SED fitting}
\label{app:sed}

\begin{figure}[]
\centering
\begin{tabular}{cc}
\includegraphics[width=0.9\columnwidth] {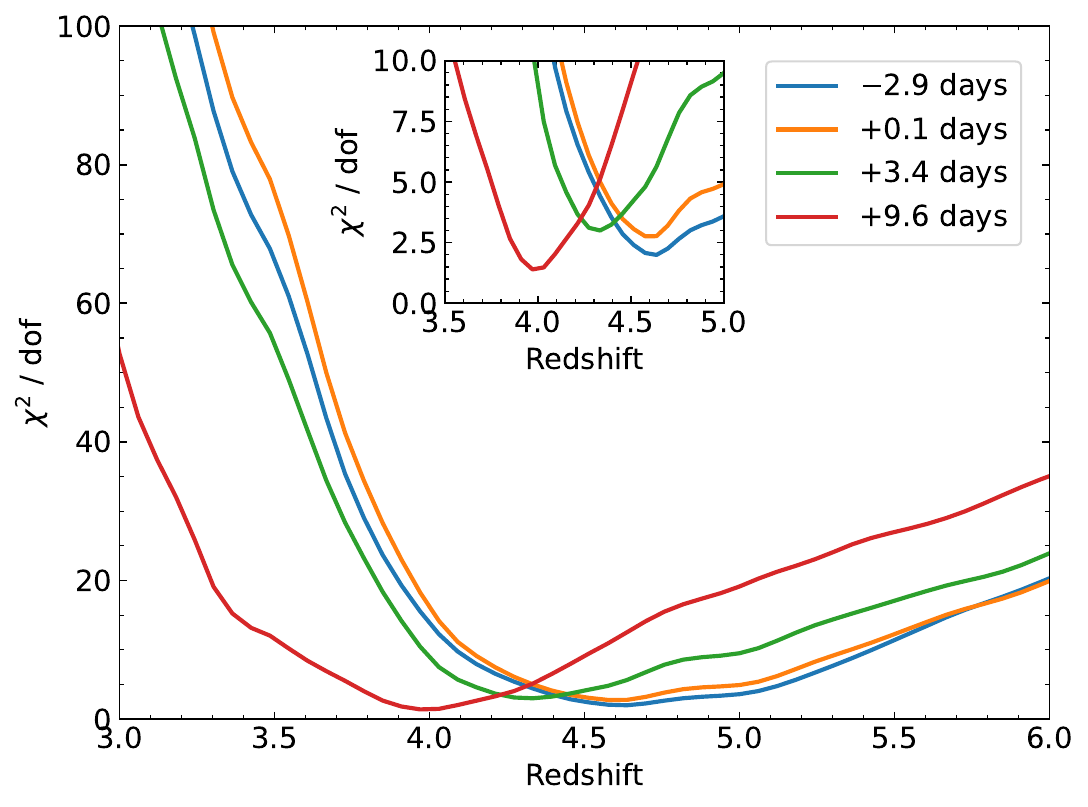} &
\end{tabular}    
	\caption{$\chi^2$/dof of the SED fitting using the spectra of SN 2011fe at different redshifts. The blue, orange, green, and red curves show the results with the spectra at -2.9, +0.1, +3.4, and +9.6 days from the maximum brightness.
    }
\label{fig:chi2_redshift}
\end{figure}

In Section \ref{sec:sed}, we perform SED fitting with spectral templates of a SN Ia (SN 2011fe).
The best-fit redshift depends on the epoch of the 
spectral template.
Figure \ref{fig:chi2_redshift} shows $\chi^2$ values of the fitting as a function of redshift for different epochs.
For the epochs around maximum ($+0.1$ days) or before the maximum ($-2.9$ days), the preferred redshift is $z \simeq 4.6$.
If we adopt a later epoch, the preferred redshift becomes smaller, as the wavelength of the spectral break evolves toward longer wavelength.
For the spectral template at $+3.4$ days and $9.6$ days, the best-fit redshifts are $z \simeq 4.3$ and $z \simeq 4.0$, respectively.

\section{Host galaxy}
\label{app:host}

We derive an upper limit on the stellar mass of the host galaxy using the off-epoch imaging, together with the non-detection at wavelengths shorter than $1.5\,\mu$m in the on-epoch data. Because the host is undetected in all off-epoch images, the available data do not warrant fitting a large parameter space of stellar population models. Instead, we adopt a simple fiducial model in which the host is a young ($t=10$\,Myr), dust-free galaxy formed in an instantaneous burst, and derive an upper limit stellar mass. We model the stellar population using the BPASS libraries \citep{2017PASA...34...58E}.

\begin{figure}
\centering
	\includegraphics[width=0.99\columnwidth]{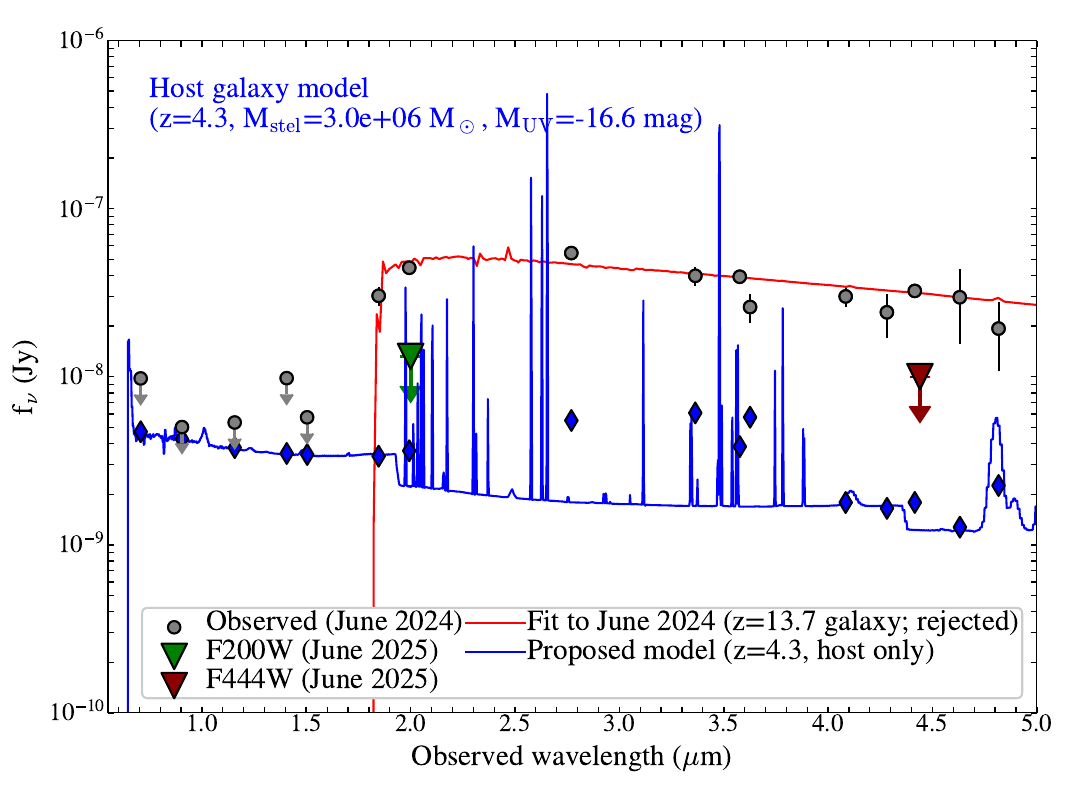}
	\caption{
    A potential host galaxy SED prediction (blue solid line; diamonds correspond to the synthetic photometric fluxes; $z=4.3, M_*=3\times10^6\,M_\odot, M_{\rm UV}=-16.6$\,mag). Data from the on-epoch (June 2024, representing the SN; gray circles) and two deepest upper limits from the off-epochs (June 2025; inverted triangles) are shown. Upper limits are $2\,\sigma$. The $z=13.7$ galaxy model \citep{2026arXiv260102861Z}, fit to the June 2024 data, is also shown (red solid line).
    }
\label{fig:sed_host}
\end{figure}

Figure~\ref{fig:sed_host} presents the upper limit model at $z=4.3$ that is consistent with all the flux upper limits, corresponding to a stellar mass of $M_*<3\times10^6\,M_\odot$ and an absolute UV magnitude of $M_{\rm UV}>-16.6$\,mag. This UV luminosity upper limit places the host among the fainter star-forming galaxies at this epoch, corresponding to $\approx30$\,th percentile of the SFR-weighted UV luminosity function, by adopting the UVLF by \citet[][]{2021AJ....162...47B}.

\section{Survey volume and duration}
\label{app:rate}

\begin{figure}[]
\centering
\includegraphics[width=0.9\columnwidth] {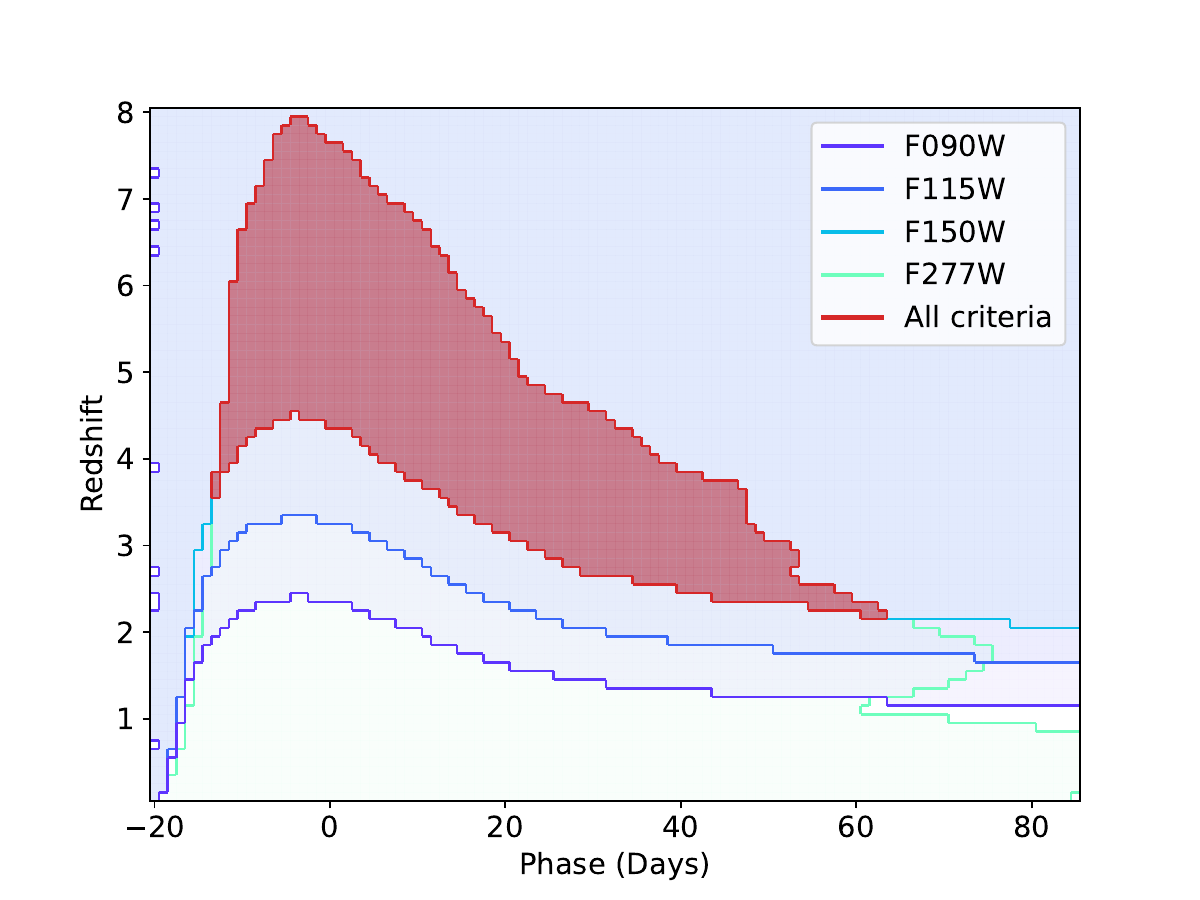}
	\caption{The allowed phase and redshift region obtained by applying high-redshift galaxy cut (${\rm SNR_{F277W}} > 4$, ${\rm SNR_{F150W}} < 2$, ${\rm SNR_{F115W}} < 2$, and  ${\rm SNR_{F090W}} < 2$) to simulated SN Ia light curves.  
    }
\label{fig:beacon-ia}
\end{figure}

Here, we describe our method for determining the survey volume and survey duration used in the rate estimation. To investigate the SN Ia survey volume and duration defined by the high-redshift galaxy classification strategy, we apply the same high-redshift galaxy selection cuts to simulated SN Ia light curves (${\rm SNR_{F277W}} > 4$, ${\rm SNR_{F150W}} < 2$, ${\rm SNR_{F115W}} < 2$, and  ${\rm SNR_{F090W}} < 2$). Although the depth varies across the BEACON field, we use the median of the limiting magnitude in the BEACON field to estimate the typical survey volume and duration for simplicity. Figure~\ref{fig:beacon-ia} shows the phase and redshift region that satisfy high-redshift galaxy selection criteria.
We use this region to evaluate the survey volume and duration (see Section \ref{sec:rate}).


\bibliography{library}{}

\bibliographystyle{aasjournalv7.1}



\end{document}

%% file: flux.tex
\begin{deluxetable}{lcc}
\tablecaption{Observed photometry of \id\ at the on-epoch.\label{tab:photometry}}
\tablehead{
\colhead{Filter} &
\colhead{$\lambda_{\rm pivot}$} &
\colhead{$f_\nu$}
}
\startdata
F070W & 7064.9 & $<9.76$ \\
F090W & 9045.3 & $<5.01$ \\
F115W & 11569.1 & $<5.34$ \\
F140M & 14059.7 & $<9.80$ \\
F150W & 15039.2 & $<5.73$ \\
F182M & 18467.5 & $30.28 \pm3.89$ \\
F200W & 19934.7 & $44.45 \pm2.60$ \\
F277W & 27698.1 & $54.36 \pm2.26$ \\
F335M & 33641.3 & $39.81 \pm5.26$ \\
F356W & 35766.4 & $39.31 \pm2.23$ \\
F360M & 36262.1 & $25.96 \pm5.25$ \\
F410M & 40842.3 & $30.01 \pm4.00$ \\
F430M & 42817.3 & $24.14 \pm7.09$ \\
F444W & 44154.4 & $32.36 \pm3.01$ \\
F460M & 46309.3 & $29.67 \pm14.04$ \\
F480M & 48167.3 & $19.32 \pm8.48$ \\
\enddata
\tablecomments{
Wavelengths are in Angstrom. Fluxes and uncertainties are in nJy. Flux errors are $1\,\sigma$. $2\,\sigma$ upper limits are shown for non-detections ($S/N<2$). 
}
\end{deluxetable}

%% file: limmag.tex
\begin{deluxetable}{cccccc}
\tablecaption{Limiting magnitudes of off-epoch imaging data}
\tablehead{
\colhead{Date} &
\colhead{$\Delta t$ (day)} &
\colhead{Filter} &
\colhead{$m_{\rm lim}$ (2$\sigma$)} &
\colhead{$t_{\rm exp}$ (s)} &
\colhead{PID}
}
\startdata
12-21-2022 & -545.81 & F115W & 29.6 & 2834.5 & 1345 \\
12-21-2022 & -545.81 & F277W & 30.4 & 2834.5 & 1345 \\
12-21-2022 & -545.74 & F356W & 30.4 & 2834.5 & 1345 \\
12-21-2022 & -545.74 & F115W & 29.6 & 2834.5 & 1345 \\
12-21-2022 & -545.70 & F150W & 29.9 & 2834.5 & 1345 \\
12-21-2022 & -545.70 & F410M & 29.6 & 2834.5 & 1345 \\
12-21-2022 & -545.65 & F444W & 29.8 & 2834.5 & 1345 \\
12-21-2022 & -545.65 & F200W & 30.1 & 2834.5 & 1345 \\
12-24-2022 & -543.03 & F115W & 29.4 & 4359.1 & 1345 \\
12-24-2022 & -543.01 & F356W & 30.0 & 1868.2 & 1345 \\
05-03-2023 & -412.19 & F182M & 29.1 & 7043.3 & 2279 \\
05-04-2023 & -412.14 & F210M & 29.1 & 8911.5 & 2279 \\
05-04-2023 & -412.10 & F444W & 29.3 & 1868.2 & 2279 \\
06-26-2023 & -358.41 & F444W & 27.8 & 4208.8 & 2234 \\
06-26-2023 & -358.41 & F090W & 30.0 & 4208.8 & 2234 \\
05-19-2025 & 333.92 & F090W & 29.5 & 3349.9 & 6434 \\
05-19-2025 & 333.97 & F200W & 29.1 & 1632.0 & 6434 \\
05-19-2025 & 333.97 & F444W & 29.5 & 1632.0 & 6434 \\
06-24-2025 & 369.95 & F200W & 30.6 & 6957.4 & 6434 \\
06-24-2025 & 369.95 & F444W & 30.4 & 6957.4 & 6434 \\
06-18-2026 & 729.68 & F162M & 29.1 & 1889.7 & 7814 \\
06-18-2026 & 729.68 & F250M & 29.5 & 3779.3 & 7814 \\
06-18-2026 & 729.71 & F140M & 29.0 & 1889.7 & 7814 \\
06-18-2026 & 729.74 & F182M & 29.4 & 1889.7 & 7814 \\
06-18-2026 & 729.74 & F460M & 28.8 & 3779.3 & 7814 \\
06-18-2026 & 729.77 & F210M & 29.3 & 1889.7 & 7814
\enddata
\tablecomments{Limiting magnitudes are measured within a $r=0.\!''16$ aperture and quoted at the 2$\sigma$ level. $\Delta t$ is the difference from the SN detection, MJD=60480.18.}
\end{deluxetable}\label{tab:data}